# Eikonal methods applied to gravitational scattering amplitudes

Stephen G. Naculich[1,a] and Howard J. Schnitzer[2,b]

[a]*Department of Physics*
*Bowdoin College, Brunswick, ME 04011, USA*

[b]*Theoretical Physics Group*
*Martin Fisher School of Physics*
*Brandeis University, Waltham, MA 02454, USA*

**Abstract**

We apply factorization and eikonal methods from gauge theories to scattering amplitudes in gravity. We hypothesize that these amplitudes factor into an IR-divergent soft function and an IR-finite hard function, with the former given by the expectation value of a product of gravitational Wilson line operators. Using this approach, we show that the IR-divergent part of the $n$-graviton scattering amplitude is given by the exponential of the one-loop IR divergence, as originally discovered by Weinberg, with no additional subleading IR-divergent contributions in dimensional regularization.

[1]Research supported in part by the NSF under grant PHY-0756518
[2]Research supported in part by the DOE under grant DE–FG02–92ER40706
naculich@bowdoin.edu, schnitzr@brandeis.edu

# 1 Introduction

Infrared divergences afflict loop-level scattering amplitudes in both gauge and gravity theories. These IR divergences (as well as any UV divergences) may be regulated using dimensional regularization in $d = 4 - 2\epsilon$ dimensions. In renormalizable theories, the UV divergences are removed from the renormalized amplitudes. The remaining IR divergences, which result from the exchange of soft virtual gauge bosons or gravitons, cancel against IR divergences arising from the emission of an arbitrary number of soft gauge bosons [1–3] or gravitons [4,5] in inclusive cross-sections.

In the scattering amplitudes of electrons or quarks, the leading IR divergence at $L$ loops is proportional to $1/\epsilon^L$. In the scattering amplitudes of massless partons, such as gluons (or massless quarks), the IR singularities are more severe, including both soft and collinear (or mass) divergences. In this case, the leading IR divergence is proportional to $1/\epsilon^{2L}$. In graviton scattering amplitudes, however, potential collinear divergences cancel from amplitudes due to momentum conservation [4] so that the leading IR divergence remains proportional to $1/\epsilon^L$ [6].

IR divergences exponentiate in both gauge [7] and gravity [4] theories, so that the leading $L$-loop IR divergence is expressed in terms of the leading IR divergence at one loop. Subleading IR divergences in gauge theories have a more complicated structure, but decades of study have produced a detailed understanding [8–21] which continues to evolve [22–30]. The IR divergences of color-singlet form factors involving massless quarks or gluons are completely characterized by the coefficients of the beta function together with two constants at each loop order, the cusp $\gamma^{(L)}$ and collinear $\mathcal{G}_0^{(L)}$ anomalous dimensions [17]. The IR divergences of scattering amplitudes involve in addition a soft anomalous dimension matrix $\Gamma^{(L)}$ (in color space [18]) at each loop order [20, 21].

The proof of the UV-finiteness of $\mathcal{N} = 8$ supergravity amplitudes through three loops [31] has renewed interest in the detailed structure of gravity scattering amplitudes. $\mathcal{N} = 8$ supergravity has now been shown to be UV-finite through six loops, based on the assumption of an $E_{7(7)}$ symmetry [32], and the situation for $L \geq 7$ has been discussed, for example, in refs. [32–39]. Using unitarity techniques, scattering amplitudes in $\mathcal{N} = 8$ supergravity can be expressed in terms of a basis of scalar integrals [31, 34, 40]. Using explicit Laurent expansions for these scalar integrals [41, 42] the two-loop four-graviton scattering amplitude was computed in refs. [43, 44]. That result revealed the absence of subleading, i.e., $\mathcal{O}(1/\epsilon)$, IR divergences at two loops in four-graviton amplitudes in $\mathcal{N} = 8$ supergravity. It was conjectured in ref. [43] that subleading IR divergences also vanish for $n$-graviton amplitudes ($n > 4$) at two loops, and possibly at higher loops as well.

In fact, Weinberg's classic paper [4] showed that one-loop IR divergences of graviton amplitudes exponentiate to all loop orders, and argued that no additional IR divergences can arise from higher-loop diagrams because soft gravitons couple with a factor of the energy of the graviton. Weinberg regulated his amplitudes using UV and IR cutoffs rather than dimensional regularization (which had not yet been invented), but his arguments would imply the absence of subleading (in the $\epsilon$-expansion) IR divergences.

In this paper, we extend the factorization approach which has been used so successfully in gauge theories [21] to gravity theories, confirming Weinberg's results, and setting the stage for corrections to the eikonal limit. We assume that $n$-graviton scattering amplitudes can be factored into a product of IR-divergent and IR-finite functions, and that the (helicity-independent) IR-divergent function can be expressed in terms of a set of eikonal loop diagrams encoded in the expectation value of a product of gravitational Wilson line operators. This approach is then used to show explicitly that potential subleading IR divergences indeed vanish. First, we reproduce the known one-loop IR divergence of the $n$-graviton amplitude in dimensional regularization [6]. We then show that the two-loop IR divergence is entirely determined by the one-loop IR divergence, with no subleading IR divergences. In particular, various two-loop diagrams, which in gauge theory do contribute to the two-loop cusp anomalous dimension, have vanishing IR-divergent contributions in gravity. This result relies on a dimensional argument, which can be extended to higher loops. We then show (heuristically) that higher-loop diagrams, other than those contributing to the exponential of the one-loop IR divergence, will be IR finite, and will not result in subleading IR divergences in the dimensionally-regulated $n$-graviton scattering amplitude. In addition to confirming this perhaps to-have-been-expected result, our methods demonstrate yet another parallel between gauge and gravity amplitudes, which have been of much recent interest.

This paper is organized as follows. In section 2, we express gravity scattering amplitudes in a factorized form, and in section 3, we hypothesize that the IR-divergent part can be written in terms of a product of gravitational Wilson lines, or equivalently, a set of eikonal loop diagrams. In sections 4 and 5, we compute the IR-divergent contributions of the one- and two-loop eikonal diagrams respectively. In section 6, we argue for the absence of subleading IR divergences at higher loops, and in section 7, we present our conclusions.

**Note added:** After the first version of this paper appeared on the arXiv, White [45] applied path integral resummation techniques from gauge theories [46] to gravity amplitudes, finding strong supporting evidence for the factorization into soft and hard functions, and confirming the form of the soft function as a product of gravitational Wilson lines hypothesized below in eq. (3.3). He also extended this approach to massive scalar external particles, as well as to next-to-eikonal order.

## 2 Infrared divergences of gravity amplitudes

Gauge theory scattering amplitudes factorize into a product of jet, soft, and hard functions (see, e.g., ref. [21]). These separate functions describe, respectively, the contributions of gluons collinear to the external lines, soft gluons exchanged between external lines, and the short-distance (infrared-finite) scattering process.

We will assume that, up to the loop order to which they are UV-finite, gravity scattering amplitudes obey a similar factorization, except that a jet function is not required since collinear divergences are absent in gravity [4]. We write the $n$-graviton scattering amplitude as

$$A_n = S_n \cdot H_n \tag{2.1}$$

where $S_n$ is an IR-divergent factor describing the exchange of soft gravitons between the $n$ external lines, and $H_n$ is IR-finite. Expanding each of the quantities in this equation in a loop expansion in powers of $\kappa^2 = 32\pi G$,

$$A_n = \sum_{L=0}^{\infty} A_n^{(L)}, \qquad S_n = 1 + \sum_{L=1}^{\infty} S_n^{(L)}, \qquad H_n = A_n^{(0)} + \sum_{L=1}^{\infty} H_n^{(L)} \qquad (2.2)$$

we have

$$A_n^{(1)} = S_n^{(1)} A_n^{(0)} + H_n^{(1)}, \qquad A_n^{(2)} = S_n^{(2)} A_n^{(0)} + S_n^{(1)} H_n^{(1)} + H_n^{(2)}, \qquad \cdots \qquad (2.3)$$

Exponentiation of leading gravitational IR divergences [4] suggests that we rewrite the soft function as

$$S_n = \exp\left[\sum_{L=1}^{\infty} s_n^{(L)}\right] \qquad (2.4)$$

i.e.

$$S_n^{(1)} = s_n^{(1)}, \qquad S_n^{(2)} = \frac{1}{2}\left[s_n^{(1)}\right]^2 + s_n^{(2)}, \qquad \cdots \qquad (2.5)$$

where the leading IR-divergence at $L$ loops is given by $\left[s_n^{(1)}\right]^L / L!$, and the $s_n^{(L)}$ contain subleading IR divergences (if any).

In $\mathcal{N} = 8$ supergravity, the all-loop-orders MHV four-graviton amplitude is proportional to the tree-level amplitude [40, 47]

$$A_4^{(L)} = M_4^{(L)} A_4^{(0)}, \qquad H_4^{(L)} = M_4^{(Lf)} A_4^{(0)}, \qquad (2.6)$$

where $M_4^{(L)}$ are helicity-independent functions of momenta, and $M_4^{(Lf)}$ is finite as $\epsilon \to 0$. Combining eqs. (2.3) and (2.6), one has

$$M_4^{(1)} = s_4^{(1)} + M_4^{(1f)}, \qquad M_4^{(2)} - \frac{1}{2}\left[M_4^{(1)}\right]^2 = s_4^{(2)} + M_4^{(2f)} - \frac{1}{2}\left[M_4^{(1f)}\right]^2, \qquad \cdots \qquad (2.7)$$

Both $M_4^{(1)}$ and $M_4^{(2)}$ may be expressed as scalar integrals [40] whose Laurent expansions in $\epsilon$ are known explicitly [41, 42]. (The explicit expansions for the scalar integrals contributing to higher-loop amplitudes [31, 34] are not yet known.) Using these, it was shown in refs. [43, 44] that the IR-divergent part of the two-loop four-point amplitude $M_4^{(2)}$ is given exactly by the IR-divergent part of one-half the square of the one-loop amplitude $M_4^{(1)}$; in other words, that $s_4^{(2)}$ is IR-finite.[3]

In ref. [43], we conjectured that this result holds more generally for $n$-graviton amplitudes in $\mathcal{N} = 8$ supergravity, i.e., that $s_n^{(2)}$ is IR-finite for any $n$. We also speculated that higher-loop subleading IR corrections, $s_n^{(L)}$ for $L \geq 2$, might also be IR-finite. In the remainder of this paper, we present an eikonal calculation of the soft function $S_n$ which supports these conjectures.

---

[3] Explicit expressions for the IR-finite part of the two-loop amplitude were also obtained in refs. [43, 44].

# 3 Eikonal description of the soft function

In gauge theories, the soft function can be given an operator interpretation in terms of infinite Wilson lines (see, e.g., refs. [19, 21, 22]), which can be expanded in terms of eikonal loop diagrams. Aybat, Dixon, and Sterman [21], employing results from ref. [48], used an eikonal calculation to compute the two-loop soft anomalous dimension matrix for $n$-parton scattering amplitudes in QCD. We will adopt a similar approach to compute the soft function for gravity amplitudes.

In ref. [4], Weinberg showed that the emission of a soft graviton (with momentum $k$) from an external line of an amplitude (with outgoing momentum $p$) multiplies the amplitude by the factor[4]

$$\left(-\frac{\kappa}{2}\right)\frac{p^\mu p^\nu}{p \cdot k + i\varepsilon} \tag{3.1}$$

(where $\kappa^2 = 32\pi G$), independent of the spin of the external line. The numerator of eq. (3.1) derives from the vertex for graviton emission, and the denominator from the propagator $(p+k)^2 - m^2 + i\varepsilon$ connecting the vertex to the rest of the amplitude, both in the eikonal limit $k \to 0$.

We introduce a gravitational Wilson line operator

$$\Phi_p(a,b) = P\exp\left(i\frac{\kappa}{2}\int_a^b ds\, p^\mu p^\nu h_{\mu\nu}(sp)\right) \tag{3.2}$$

where $s$ parametrizes a straight line segment $x^\mu = sp^\mu$. By Fourier-transforming the exponent of eq. (3.2) to momentum space, we see that it gives rise to a vertex in which a graviton $h_{\mu\nu}$ of momentum $k$ couples to a source (3.1). We hypothesize (in analogy with the gauge theory expressions in refs. [21, 22]) that the soft function in the factorization (2.1) may be expressed in terms of a product of infinite gravitational Wilson lines

$$S_n = \left\langle 0 \left| \prod_{i=1}^n \Phi_{p_i}(0,\infty) \right| 0 \right\rangle. \tag{3.3}$$

The expansion of $S_n$ in powers of $\kappa$ then yields a series of eikonal loop diagrams in which soft gravitons are exchanged between $n$ external lines of momenta $p_i$. By our hypothesis, the IR divergences of these eikonal diagrams reproduce the IR divergences (both leading and subleading) of the original amplitude, as is the case for gauge theory.

We observe that eq. (3.3) is the specialization (up to normalization) of a more general Wilson loop operator introduced by Brandhuber, Heslop, Nasti, Spence, and Travaglini [44] (see also ref. [49])

$$W[\mathcal{C}] = \left\langle 0 \left| P\exp\left(i\kappa \oint_\mathcal{C} d\tau\, \dot{x}^\mu(\tau)\dot{x}^\nu(\tau) h_{\mu\nu}(x(\tau))\right) \right| 0 \right\rangle \tag{3.4}$$

where $x^\mu(\tau)$ parametrizes a contour $\mathcal{C}$. These authors explored whether MHV amplitudes in $\mathcal{N} = 8$ supergravity could be expressed in terms of gravitational Wilson loop expectation values.

---

[4]In this paper, we use the mostly-minus metric, $\eta_{00} = 1$, whereas Weinberg uses the mostly-plus metric.

(This was motivated by the discovery that the planar MHV $n$-point amplitude in $\mathcal{N} = 4$ SYM theory, both IR-divergent and IR-finite parts, could be calculated in terms of expectation values of polygonal Wilson loops [50–56].) As observed in ref. [44], however, the gravitational Wilson loop (3.4), unlike the gauge theory Wilson loop, is not in general invariant under

$$h_{\mu\nu} \to h_{\mu\nu} + \partial_\mu \xi_\nu + \partial_\nu \xi_\mu \,. \tag{3.5}$$

As was further observed in ref. [44], if the contour $\mathcal{C}$ is composed of straight segments, the non-invariance of eq. (3.4) is concentrated at the endpoints of the segments. The cusps of the polygons result in gauge-variant expressions for polygonal gravitational Wilson loops, casting doubt on this prescription (although for a certain choice of gauge, the resulting expressions can be made to match the one-loop four-graviton scattering amplitude).

In contrast, the variation of eq. (3.3) under eq. (3.5) vanishes because of momentum conservation

$$\sum_{i=1}^{n} p_i^\mu = 0 \tag{3.6}$$

provided that $\xi_\mu(\infty)$ is independent of direction. Consequently, the soft function $S_n$ is gauge-invariant.

In the following two sections, we will use eq. (3.3) to generate one- and two-loop eikonal diagrams to compute the IR divergences of $n$-graviton scattering amplitudes.

## 4  One-loop IR divergences of gravity amplitudes

Expanding the soft function (3.3) to one-loop order gives rise to a set of eikonal diagrams, in which a graviton (thin line) is exchanged between an arbitrary pair $i, j$ of external legs (thick lines), cf fig. 1.

We will find it convenient to evaluate the resulting eikonal diagrams in momentum space. Using the graviton propagator in de Donder gauge,

$$P_{\mu\nu,\alpha\beta} = \frac{i}{2} \left[ \eta_{\mu\alpha}\eta_{\nu\beta} + \eta_{\mu\beta}\eta_{\nu\alpha} - \frac{2}{d-2}\eta_{\mu\nu}\eta_{\alpha\beta} \right] \frac{1}{k^2 + i\varepsilon} \tag{4.1}$$

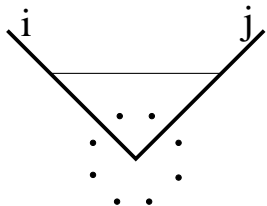

Figure 1: The one-loop eikonal diagram contributing to $S_n^{(1)}$. A graviton (thin line) is exchanged between external legs $i$ and $j$ (thick lines). The dots signify an arbitrary number of other external legs.

together with eikonal vertex (3.1), we obtain

$$S_n^{(1)} = i\left(\frac{\kappa\mu^\epsilon}{2}\right)^2 \sum_{i<j} \int \frac{d^d k}{(2\pi)^d} \frac{(2p_i \cdot p_j)^2}{(k^2 + i\varepsilon)(2k \cdot p_i + i\varepsilon)(-2k \cdot p_j + i\varepsilon)} \tag{4.2}$$

where we have replaced $\kappa \to \kappa\mu^\epsilon$ in eq. (3.1) (where $\mu$ is an arbitrary mass scale) to keep $S_n^{(1)}$ dimensionless in $d = 4 - 2\epsilon$ dimensions. We observe that the last term in the de Donder propagator (4.1) does not contribute to $S_n^{(1)}$ (nor to any of the two-loop diagrams considered in the next section) because of the masslessness of the external lines, $p_i^2 = p_j^2 = 0$.

The integral (4.2) formally vanishes due to cancellation between IR and UV divergences. The UV divergence is artificial (unrelated to any actual UV divergences of the gravity amplitudes), and results from the fact that we are using the eikonal ($k \to 0$) limit for soft-graviton exchange as if it were valid for all values of $k$. Once we remove these artificial UV divergences, $S_n^{(1)}$ will give the IR divergence of the amplitude.

We Feynman parametrize eq. (4.2) as

$$\begin{aligned} S_n^{(1)} &= 2i\left(\frac{\kappa\mu^\epsilon}{2}\right)^2 \sum_{i<j}(2p_i \cdot p_j)^2 \int \frac{d^d k}{(2\pi)^d} \int_0^\infty dx_1 \int_0^\infty dx_2 \frac{1}{[k^2 + 2k \cdot (x_1 p_i - x_2 p_j) + i\varepsilon]^3} \\ &= \left(\frac{\kappa\mu^\epsilon}{2}\right)^2 \frac{\Gamma(1+\epsilon)}{(4\pi)^{d/2}} \sum_{i<j}(2p_i \cdot p_j)^2 \int_0^\infty dx_1 \int_0^\infty dx_2 \frac{1}{[(x_1 p_i - x_2 p_j)^2]^{1+\epsilon}} \,. \end{aligned} \tag{4.3}$$

Since the external lines are massless ($p_i^2 = p_j^2 = 0$), this simplifies to

$$S_n^{(1)} = \left(\frac{\kappa\mu^\epsilon}{2}\right)^2 \frac{\Gamma(1+\epsilon)}{(4\pi)^{d/2}} \sum_{i<j}(-2p_i \cdot p_j)^{1-\epsilon} \int_0^\infty dx_1 \int_0^\infty dx_2 (x_1 x_2)^{-1-\epsilon} \,. \tag{4.4}$$

To disentangle UV and IR divergences, we follow the strategy of ref. [57] and use the symmetry under $x_1 \leftrightarrow x_2$ to rewrite this as

$$S_n^{(1)} = \left(\frac{\kappa\mu^\epsilon}{2}\right)^2 \frac{\Gamma(1+\epsilon)}{(4\pi)^{d/2}} \sum_{i<j}(-2p_i \cdot p_j)^{1-\epsilon} \int_0^\infty dx_1 \frac{1}{x_1^{1+\epsilon}} \int_0^{x_1} dx_2 \frac{2}{x_2^{1+\epsilon}} \tag{4.5}$$

and then take $\epsilon < 0$ to regulate the IR divergence at $x_2 = 0$:

$$S_n^{(1)} = \left(\frac{\kappa\mu^\epsilon}{2}\right)^2 \frac{\Gamma(1+\epsilon)}{(4\pi)^{d/2}} \sum_{i<j}(-2p_i \cdot p_j)^{1-\epsilon} \left(\frac{-2}{\epsilon}\right) \int_0^\infty dx_1 x_1^{-1-2\epsilon} \,. \tag{4.6}$$

The integral over $x_1$ formally vanishes as a result of cancellation of UV and IR poles. We cancel the UV pole by adding a counterterm $-1/(2\epsilon)$ to the integral over $x_1$, which leaves us with

$$\begin{aligned} S_n^{(1)} &= \left(\frac{\kappa\mu^\epsilon}{2}\right)^2 \frac{\Gamma(1+\epsilon)}{(4\pi)^{d/2}} \sum_{i<j} \frac{(-2p_i \cdot p_j)^{1-\epsilon}}{\epsilon^2} \\ &= \frac{\lambda}{(4\pi)^2} \sum_{i<j} \frac{(-2p_i \cdot p_j)}{\epsilon^2} \left(\frac{\mu^2}{-2p_i \cdot p_j}\right)^\epsilon + \mathcal{O}(\epsilon^0) \end{aligned} \tag{4.7}$$

where we have introduced
$$\lambda = \left(\frac{\kappa}{2}\right)^2 \left(4\pi e^{-\gamma}\right)^\epsilon. \tag{4.8}$$

The leading $1/\epsilon^2$ pole in eq. (4.7) cancels via momentum conservation $\sum_{i<j} p_i \cdot p_j = 0$ and we are left with the IR-divergent contribution to the one-loop $n$-graviton amplitude
$$S_n^{(1)} = \frac{\lambda}{(4\pi)^2 \epsilon} \sum_{i<j} (2 p_i \cdot p_j) \log\left(\frac{-2 p_i \cdot p_j}{\mu^2}\right) + \mathcal{O}(\epsilon^0). \tag{4.9}$$

This is precisely the result obtained by Dunbar and Norridge [6]. For $n = 4$, eq. (4.9) reduces to
$$S_4^{(1)} = \frac{\lambda}{8\pi^2 \epsilon} \left[ s \log\left(\frac{-s}{\mu^2}\right) + t \log\left(\frac{-t}{\mu^2}\right) + u \log\left(\frac{-u}{\mu^2}\right) \right] + \mathcal{O}(\epsilon^0) \tag{4.10}$$

(where $s$, $t$, and $u$ are the Mandelstam variables), which is indeed the IR-divergent part of the known $\mathcal{N} = 8$ four-graviton amplitude [40, 43, 44, 58].

## 5 Two-loop IR divergences of gravity amplitudes

Expanding the soft function (3.3) to two-loop order generates the same set of eikonal diagrams used in ref. [21] to compute the two-loop soft anomalous dimension matrix for gauge theories, cf fig. 2. It is again most convenient to evaluate these diagrams in momentum space.

The ladder and crossed-ladder diagrams shown in figs. 2(a) and 2(b), respectively, give rise to the contributions
$$-\left(\frac{\kappa \mu^\epsilon}{2}\right)^4 \sum_{i<j} \int \frac{d^d k_1}{(2\pi)^d} \frac{d^d k_2}{(2\pi)^d} \frac{(p_i \cdot p_j)^4}{k_1^2 k_2^2 (k_1 \cdot p_j)(k_1+k_2) \cdot p_j (k_1+k_2) \cdot p_i} \left[\frac{1}{(k_1 \cdot p_i)} + \frac{1}{(k_2 \cdot p_i)}\right]$$
$$= \frac{1}{2}\left(\frac{\kappa \mu^\epsilon}{2}\right)^4 \sum_{i<j} \left[\int \frac{d^d k}{(2\pi)^d} \frac{-i(p_i \cdot p_j)^2}{k^2 (k \cdot p_i)(k \cdot p_j)}\right]^2 \tag{5.1}$$

The diagrams involving three eikonal legs shown in figs. 2(c) and 2(d) contribute
$$-\frac{1}{2}\left(\frac{\kappa \mu^\epsilon}{2}\right)^4 \sum_{j} \sum_{i \neq j} \sum_{k \neq i,j} \int \frac{d^d k_1}{(2\pi)^d} \frac{d^d k_2}{(2\pi)^d} \frac{(p_i \cdot p_j)^2 (p_j \cdot p_k)^2}{k_1^2 k_2^2 (k_1 \cdot p_i)(k_2 \cdot p_k)(k_1+k_2) \cdot p_j} \left[\frac{1}{(k_2 \cdot p_j)} + \frac{1}{(k_1 \cdot p_j)}\right]$$
$$= \frac{1}{2}\left(\frac{\kappa \mu^\epsilon}{2}\right)^4 \sum_{j} \sum_{i \neq j} \sum_{k \neq i,j} \int \frac{d^d k_1}{(2\pi)^d} \left[\frac{-i(p_i \cdot p_j)^2}{k_1^2 (k_1 \cdot p_i)(k_1 \cdot p_j)}\right] \int \frac{d^d k_2}{(2\pi)^d} \left[\frac{-i(p_j \cdot p_k)^2}{k_2^2 (k_2 \cdot p_j)(k_2 \cdot p_k)}\right] \tag{5.2}$$

The diagram involving four eikonal legs in fig. 2(e) contributes
$$\frac{1}{2}\left(\frac{\kappa \mu^\epsilon}{2}\right)^4 \sum_{i<j} \sum_{\substack{k<l \\ k \neq i,j \\ l \neq i,j}} \int \frac{d^d k_1}{(2\pi)^d} \left[\frac{-i(p_i \cdot p_j)^2}{k_1^2 (k_1 \cdot p_i)(k_1 \cdot p_j)}\right] \int \frac{d^d k_2}{(2\pi)^d} \left[\frac{-i(p_k \cdot p_l)^2}{k_2^2 (k_2 \cdot p_k)(k_2 \cdot p_l)}\right] \tag{5.3}$$

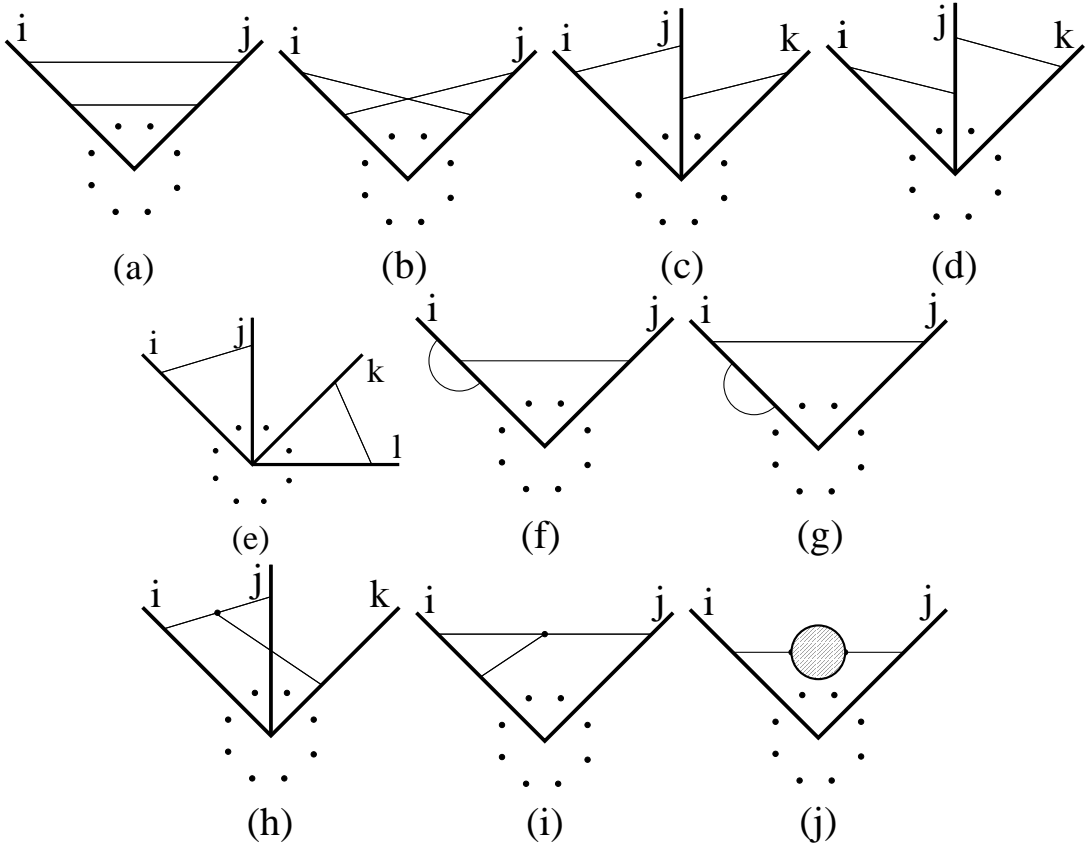

Figure 2: Two-loop eikonal diagrams contributing to $S_n^{(2)}$

The factors of $\frac{1}{2}$ in eqs. (5.2) and (5.3) compensate for double counting some of the diagrams. The sum of the Abelian-like contributions (5.1)-(5.3) can be neatly rewritten as one-half the square of the one-loop eikonal contribution (4.2), namely

$$\frac{1}{2}\left[S_n^{(1)}\right]^2 = \frac{1}{2}\left[\left(\frac{\kappa\mu^\epsilon}{2}\right)^2 \sum_{i<j} \int \frac{d^dk}{(2\pi)^d} \frac{-i(p_i\cdot p_j)^2}{k^2(k\cdot p_i)(k\cdot p_j)}\right]^2 \qquad (5.4)$$

as was observed by Weinberg long ago [4].

The 1PI diagrams shown in figs. 2(f) and 2(g), in which one graviton is emitted and absorbed by the same eikonal line, contain factors of $p_i^2$ in their numerators, and hence vanish for massless external lines.

Figure 2(h) involves a triple graviton vertex

$$i\left(\frac{\kappa\mu^\epsilon}{2}\right)^3 \sum_{i<j<k} \int \frac{d^dk_1}{(2\pi)^d}\frac{d^dk_2}{(2\pi)^d}\frac{d^dk_3}{(2\pi)^d} \frac{\delta^{(d)}(k_1+k_2+k_3)}{k_1^2\,k_2^2\,k_3^2\,(k_1\cdot p_i)(k_2\cdot p_j)(k_3\cdot p_k)} V_3(k_1,k_2,k_3;p_i,p_j,p_k) \qquad (5.5)$$

where

$$V_3(k_1,k_2,k_3;p_i,p_j,p_k) = p_i^\mu p_i^\nu p_j^\alpha p_j^\beta p_k^\sigma p_k^\tau G_{\mu\nu,\alpha\beta,\sigma\tau}(k_1,k_2,k_3) \qquad (5.6)$$

is given by the numerators from the gravitational eikonal vertices contracted with the three-graviton vertex $G_{\mu\nu,\alpha\beta,\sigma\tau}(k_1, k_2, k_3)$. This vertex is in general gauge-dependent. Using the expression obtained by Bern and Grant[5] one has [59]

$$V_3(k_1, k_2, k_3; p_i, p_j, p_k) = -i\left(\frac{\kappa\mu^\epsilon}{2}\right)\left\{\left[(p_i \cdot p_j)(p_k \cdot k_1) + (p_j \cdot p_k)(p_i \cdot k_2) + (p_k \cdot p_i)(p_j \cdot k_3)\right]^2 \right.$$
$$\left. + \left[(p_i \cdot p_j)(p_k \cdot k_2) + (p_j \cdot p_k)(p_i \cdot k_3) + (p_k \cdot p_i)(p_j \cdot k_1)\right]^2\right\} \quad (5.7)$$

which is manifestly even under $k_i \to -k_i$. The three-graviton vertex in other gauges is more complicated (see e.g. ref. [5]) but shares the property of being even under $k_i \to -k_i$. This implies that the integrand in eq. (5.5) is odd under $k_i \to -k_i$, hence the integral vanishes independent of gauge. It therefore has no UV divergence, and no counterterm, so this diagram does not contribute to the soft function. (In nonabelian gauge theories, the analogous diagram also vanishes [21].)

Figure 2(i), which contributes

$$i\left(\frac{\kappa\mu^\epsilon}{2}\right)^3 \sum_{i<j} \int \frac{d^d k_1}{(2\pi)^d} \frac{d^d k_2}{(2\pi)^d} \frac{d^d k_3}{(2\pi)^d} \frac{\delta^{(d)}(k_1 + k_2 + k_3)}{k_1^2 \, k_2^2 \, k_3^2 \, (k_1 \cdot p_i)(k_3 \cdot p_i)(k_3 \cdot p_j)} V_3(k_1, k_2, k_3; p_i, p_i, p_j) \quad (5.8)$$

also vanishes because the integrand is odd under $k_i \to -k_i$. Hence, this diagram does not contribute to the soft function either. (In nonabelian gauge theories, the analogous diagram does contribute [21].)

The final diagram, shown in fig. 2(j), contains a self-energy insertion in the graviton propagator exchanged between two eikonal legs, and contributes

$$i\left(\frac{\kappa\mu^\epsilon}{2}\right)^2 \sum_{i<j} \int \frac{d^d k}{(2\pi)^d} \frac{p_i^\mu p_i^\nu p_j^\alpha p_j^\beta \Pi_{\mu\nu,\alpha\beta}(k)}{k \cdot p_i \, (k^2)^2 \, k \cdot p_j} \quad (5.9)$$

where the one-loop graviton polarization tensor has the form [60]

$$\Pi^{\mu\nu,\alpha\beta}(k) = \left(\frac{\kappa\mu^\epsilon}{2}\right)^2 \left[k^\mu k^\nu k^\alpha k^\beta \Pi_1(k^2) + k^2(k^\mu k^\nu \eta^{\alpha\beta} + k^\alpha k^\beta \eta^{\mu\nu})\Pi_2(k^2)\right.$$
$$+ k^2(k^\mu k^\alpha \eta^{\nu\beta} + k^\mu k^\beta \eta^{\nu\alpha} + k^\nu k^\alpha \eta^{\mu\beta} + k^\nu k^\beta \eta^{\mu\alpha})\Pi_3(k^2)$$
$$\left. + (k^2)^2 \eta^{\mu\nu}\eta^{\alpha\beta}\Pi_4(k^2) + (k^2)^2(\eta^{\mu\alpha}\eta^{\nu\beta} + \eta^{\mu\beta}\eta^{\nu\alpha})\Pi_5(k^2)\right]. \quad (5.10)$$

---

[5]The Bern-Grant gauge was developed for calculating tree-level graviton scattering, and has not been validated for loop calculations. However, use of the Bern-Grant gauge does not alter any of the eikonal results computed in this paper provided the external lines are massless.

Contractions with the external momenta in eq. (5.9) leaves

$$i\left(\frac{\kappa\mu^\epsilon}{2}\right)^4 \sum_{i<j}\int\frac{d^dk}{(2\pi)^d}\left[\frac{(k\cdot p_i)(k\cdot p_j)}{(k^2)^2}\Pi_1(k^2) + \frac{4(p_i\cdot p_j)}{k^2}\Pi_3(k^2) + \frac{2(p_i\cdot p_j)^2}{(k\cdot p_i)(k\cdot p_j)}\Pi_5(k^2)\right]. \quad (5.11)$$

Assuming the fields contributing to the graviton self-energy are massless, the $\Pi_i(k^2)$ all have momentum dependence $(k^2)^{-\epsilon}$, so the integrands of each of the terms in eq. (5.11) go as $(k^2)^{-1-\epsilon}$, and therefore the integral does not diverge in the IR region $k\to 0$. Consequently, fig. 2(j) can only give an IR-finite contribution to the $n$-graviton amplitude.

To summarize, the five two-loop eikonal diagrams fig. 2(a)-(e) add up to give precisely one-half the square of the one-loop eikonal diagram. None of the remaining five diagrams fig. 2(f)-(j) contain IR-divergent contributions. Therefore we have

$$S_n^{(2)} = \frac{1}{2}\left[S_n^{(1)}\right]^2 + \text{(IR finite)} \qquad \Longrightarrow \qquad s_n^{(2)} = \text{(IR finite)} \quad (5.12)$$

i.e., there are no subleading IR-divergences in the two-loop $n$-graviton scattering amplitude. This confirms the explicit finding for the two-loop four-graviton amplitude in $\mathcal{N}=8$ supergravity [43, 44], and extends it to the $n$-graviton amplitude.

# 6  Higher-loop IR divergences of gravity amplitudes

In the previous section, we computed the eikonal contribution to the two-loop $n$-graviton scattering amplitude, and explicitly showed the absence of subleading IR divergences. These calculations, and in particular that of the self-energy diagram, suggest a strategy for showing the absence of subleading IR corrections at arbitrary loop order. While we make no claim to rigor, the arguments of this section suggest that the only diagrams contributing to the IR-divergent part of the amplitude are those generated by the expansion of the exponential of the one-loop diagram.

Consider an $L$-loop $n$-graviton eikonal diagram containing $E$ eikonal vertices at which a soft-graviton attaches to an external line. Schematically such a diagram gives a contribution

$$\prod_{j=1}^L \int d^d\ell_j\, f_{\mu_1\nu_1\cdots\mu_E\nu_E}(\ell) \prod_{a=1}^E \left(-\frac{\kappa\mu^\epsilon}{2}\right)\frac{p_a^{\mu_a}p_a^{\nu_a}}{p_a\cdot k_a}. \quad (6.1)$$

Here, $p_a$ is the momentum of the external line from which the soft graviton is emitted; the $p_a$ are not necessarily distinct. The momentum of the emitted graviton, $k_a$, is some linear combination of the independent loop momenta $\ell_j$. The function $f_{\mu_1\nu_1\cdots\mu_E\nu_E}(\ell)$ is composed of internal propagators and vertices. Overall, the internal vertices contribute a factor of $(\kappa\mu^\epsilon)^{2L-E}$ in order that the whole diagram goes as $(\kappa\mu^\epsilon)^{2L}$. Assuming that all the interacting fields are massless, $f_{\mu_1\nu_1\cdots\mu_E\nu_E}(\ell)$ is homogeneous of degree $-2L$ in the loop momenta $\ell$ to make the final result dimensionless. Hence, the integral scales as

$$(\kappa\mu^\epsilon)^{2L}p^E \int \frac{d^{dL}\ell}{\ell^{2L+E}} \quad (6.2)$$

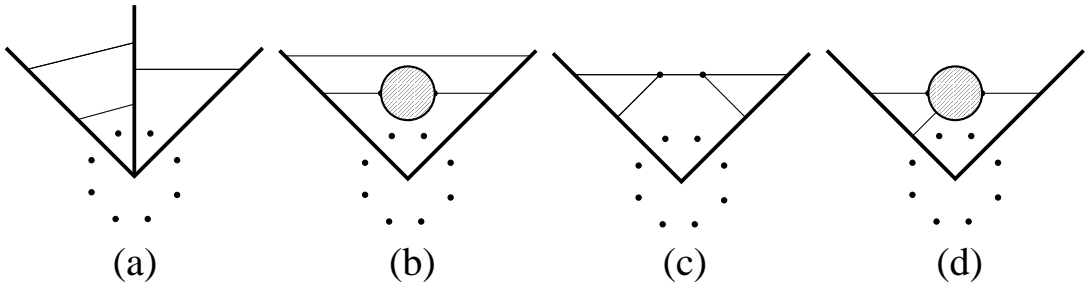

Figure 3: Some three-loop eikonal diagrams

At this point, we divide the $L$-loop eikonal diagrams into two classes: those with $E = 2L$ and those with $E < 2L$. The former class involves independent exchanges of $L$ soft gravitons between external lines, e.g. figs. 2(a)-(e) of the previous section, or fig. 3(a). The integrals over loop momenta in eq. (6.2) generate IR divergences in these diagrams. By a generalization of the calculation of the previous section, however, the sum of all such Abelian-like diagrams is given by [4]

$$\frac{1}{L!} \left[ S_n^{(1)} \right]^L . \tag{6.3}$$

(Diagrams in the class $E = 2L$ in which a soft graviton is emitted and absorbed by the same external line, e.g. figs. 2(f) and (g), vanish before integration because of the presence of factors of $p^2$ in the integrand.)

The diagrams in the class $E < 2L$ all contain non-trivial internal vertices, e.g. figs. 3(b)-(d). Diagrams like fig. 3(b) also have IR divergences, but these are generated from the IR divergences of lower loop diagrams. For example fig. 3(b) contributes to the term $s_n^{(1)} s_n^{(2)}$ in the expansion of $S_n$ in eq. (2.4).

Finally, diagrams like fig. 3(c) and (d), in which all the soft gravitons are connected in a connected web [61, 62], are not generated by lower-loop diagrams, and contribute directly to $s_n^{(3)}$. One may characterize such diagrams as non-Abelian-like. Included in this category are diagrams with vertices containing four (or more) gravitons. If the loop momenta in these diagrams are all scaled uniformly, power counting suggests that eq. (6.2) has no IR divergence as $\ell \to 0$. This suggests that the functions $s_n^{(L)}$ are all IR-finite. Consequently, subleading IR divergences are absent in the $n$-graviton amplitude.

# 7 Conclusions

In this paper, we have extended to gravity the powerful factorization methods used in gauge theory. These methods imply that scattering amplitudes can be factored into an IR-divergent soft function and an IR-finite hard function, and that the soft function can be written as the vacuum expectation value of a product of gravitational Wilson loops. Using this approach, we calculated the IR divergences of $n$-graviton scattering amplitudes, and confirmed the classic result of Weinberg [4] that these are generated by the exponential of the one-loop IR divergences, with no additional IR-divergent contributions.

There are various open issues which deserve mention:

(1) In this paper we assumed that the factorization of amplitudes is valid, at least to the order in perturbation theory to which the gravity theory is UV finite. In ref. [45], strong supporting evidence for this hypothesis has been presented.

(2) What happens at the loop order at which the gravity theory is UV-divergent, and thus requires UV subtractions? Does the factorization fail, or do the regulated UV divergences and accompanying undetermined parameters become part of the definition of the hard function, with factorization intact?

(3) What happens for massive external Wilson lines? White [45] has now studied this generalization.

It should also be mentioned that eikonal resummation in gravity and supergravity has been used in ref. [63] to study high-energy scattering in the ultra-Planckian limit. Although there does not appear to be a direct connection with our use of the eikonal approximation to determine IR divergences, this issue deserves further consideration.

## Acknowledgments

We would particularly like to thank J. Henn for his initial participation in this project, and for helpful comments throughout. We would also like to acknowledge useful conversations and correspondence with L. Dixon and H. Elvang. Finally, we thank C. White for calling our attention to his recent work.